# Ship resistance when operating in floating ice floes: a combined CFD&DEM approach

Luofeng Huang[a, *], Jukka Tuhkuri[b], Bojan Igrec[a], Minghao Li[c], Dimitris Stagonas[a, d],

Alessandro Toffoli[e], Philip Cardiff[f], Giles Thomas[a]

[a]Department of Mechanical Engineering, University College London, United Kingdom
[b]Department of Mechanical Engineering, Aalto University, Finland
[c]Department of Mechanics and Maritime Science, Chalmers University of Technology, Sweden
[d]Department of Offshore Process and Energy Engineering, Cranfield University, United Kingdom
[e]Department of Infrastructure Engineering, University of Melbourne, Australia
[f]School of Mechanical and Materials Engineering, University College Dublin, Ireland

**Abstract**

Whilst climate change is transforming the Arctic into a navigable ocean where small ice floes are floating on the sea surface, the effect of such ice conditions on ship performance has yet to be understood. The present work combines a set of numerical methods to simulate the ship-wave-ice interaction in such ice conditions. Particularly, Computational Fluid Dynamics is applied to provide fluid solutions for the floes and it is incorporated with the Discrete Element Method to govern ice motions and account for ship-ice/ice-ice collisions, by which, the proposed approach innovatively includes wave effects in the interaction. In addition, this work introduces two algorithms that can implement computational models with natural ice-floe fields, which takes randomness into consideration thus achieving high-fidelity modelling of the problem. Following validation against experiments, the model is shown accurate in predicting the ice-floe resistance of a ship, and then a series of simulations are performed to investigate how the resistance is influenced by ship speed, ice concentration, ice thickness and floe diameter. This paper presents a useful approach that can provide power estimates for Arctic shipping and has the potential to facilitate other polar engineering purposes.

**Keywords:** Ice floe; Ship Resistance; Computational Fluid Dynamics; Discrete Element Method.

# 1. Introduction

Global warming has induced a paradigm change in the Arctic environment, pronounced by an obvious transition from level-ice coverages to broken ice-floe fields and open water [1]. The changed condition makes the region more accessible to ships, with numerous waterways opening for travelling between continents and the Arctic, which are used to access oil, gas, mines, fishing grounds and tourism. In addition, there are two major cargo-shipping routes becoming available, the Northwest Passage and the Northern Sea Route, which can be used as alternatives to the Panama and Suez canals to connect Europe, Asia and America; compared with their current counterparts, both new routes can reduce the travel distance by up to 40%, leading to significant fuel, cost, time and emission savings [2]. Under this trend, Arctic shipping is now attracting significant investment from commercial stakeholders and of special research interest.

There are formidable challenges coming hand-in-hand with the benefits of Arctic shipping. One of the most obvious is to understand the effect of the potential navigation environment on ships: instead of providing pure open water, level ice coverages are broken up into numerous ice floes floating on the sea surface. These ice floes, also known as pancake ice, tend to be circular under the effect of wave wash and floe–floe collisions, as shown in Figure 1. Pancake ice has been predicted to be the most ubiquitous condition for future Arctic shipping [1], and it is also dominant in the Antarctic [3]. This has motivated experiments on ship advancement in ice floes [4–7], which reported the floes can induce significant resistance increments on the ship, indicating it is of great importance to predict the ice-added resistance. Nevertheless, considering the prohibitive costs of experiments and the shortage of field-measurement data, developing a reliable computational model to predict the ship performance with ice floes can be a cost-effective way to provide convenient solutions.

Successful modelling of broken ice-floe fields has been achieved using the Discrete Element Method (DEM), since this method allows the calculation of the contact force of ice-ice and ice-structure, which is essential for engineering problems in such ice-floe fields. A review on this has been given by Tuhkuri and Polojarvi [8]. Using DEM, Løset [9,10] calculated the force of an array of ice floes on a beam and found the load increases with an increased ice concentration, ice drift speed and the width of the beam, in which, the load appears to vary linearly with drift speed and beam width, while exponentially with ice concentration. Hopkins and Tuhkuri [11] simulated the jam of pancake ice floes in a trench and provided parallel experimental comparisons to show the overturning and rafting of floes can be decently modelled by DEM. Hopkins and Shen [12] further applied the method to replicate pancake-ice rafting in waves and the model was subsequently validated by the wave-tank experiments of Dai et al. [13]. Herman et al. [14–16] conducted combined numerical-experimental studies on wave attenuation through ice-floe fields, with DEM to account for the energy dissipation due to floe collisions. Moreover, Gong et al. [17] studied the resistance of unconsolidated ice ridges on a ship, where the resistance was reported partially due to ice friction and partially attributed to pushing the ice blocks; particularly, the second component varies linearly with the mass of the ice pushed by the ship.

Currently, one gap in related DEM simulations is how to accurately account for the force of the surrounding fluid on ice, which is usually implemented by empirical equations [8]. Due to this deficiency, previous simulations of a ship advancing in ice floes ignored the effect of fluid flow [7,18–20], which can make the modelling not realistic enough. The process of a ship advancing in floating ice floes can be summarised as the following ship-wave-ice interaction: ship advancement generates waves; waves interact with ice floes; ice floes contact each other and with the ship. The ship-generated waves can play a key role within the process; for example, it can change the velocity (magnitude and direction) of floes, especially when the floes are small. Therefore, ignoring the wave effect may considerably influence the ice load on a ship.

Tsarau et al. [21,22] incorporated the potential-flow theory with DEM to better account for the fluid force on ice; however, recent experiments [23–25] have demonstrated that the linear assumptions applied in the potential-flow theory can cause inaccuracies in certain scenarios. For example, the theory excludes a phenomenon, overwash, defined as waves running over the top surface of floating ice [26], which can evidently suppress the movement of floes [24]. Since sea ice has very small freeboard due to its similar density with water, overwash is highly frequent in ship-wave-ice interactions. Thus, the exclusion of overwash makes the potential-flow theory less applicable to the proposed problem.

An alternative approach is to couple Computational Fluid Dynamics (CFD) with DEM, which allows for fully non-linear fluid solutions and is one of the most mature approaches to model hydrodynamic problems [27]. CFD has been shown capable of simulating the motion of an ice floe in waves and has achieved good agreement with experiments when overwash occurs [28–30]. In addition, CFD is convenient for including complex geometries to study wave-ice-structure interactions, and it has been proved to accurately obtain the wave generated by an advancing ship [31]. Another alternative has been found coupling the Lattice Boltzmann method (LBM) with structural solutions to model the ship-wave-ice interaction [32], but the application of LBM in ocean engineering is still rare and might require more validations. One disadvantage of CFD comparing with theoretical predictions is that it needs much higher computational power. However, as DEM requests very high computational power itself, compared with using DEM solitarily, using CFD to provide fluid solutions for DEM will not significantly increase the required computational cost. Based on the above reasons, this study proposes to combine CFD with DEM to achieve the ship-wave-ice coupling.

Apart from accurately accounting for the fluid force, another challenge in the modelling of ice-floe fields is to import natural ice distributions into computational models. In polar regions, ice floes are randomly distributed and of a range of sizes [3,33]. Even though computational models are capable of simulating the structure-wave-ice interaction, the initial size and location of each floe need to be prescribed. For example, in the computational models of Janssen et al. [32], Sun and Shen [34] and Huang et al. [35], ice floes are set to be of a uniform size and the distance between all ice floes is initialised to be the same, as shown in Figure 2; this is not a natural condition and the result can be subjected to initial setups (e.g. the relative positions between ship and floes considerably influence the ice resistance [35]). Therefore, there is the requirement to import natural ice distributions to realistically simulate the physical processes associated with floe fields.

The present paper presents a set of methods for modelling natural ice-floe fields and the simulation of a ship advancing in such an ice condition; subsequently, it focusses on investigating the ice-added resistance on the ship. Section 2.1 and 2.2 demonstrate the CFD and DEM methods applied in the proposed problem and how they are coupled together to model the ship-wave-ice interaction; Section 2.3 presents the development of two algorithms that can generate ice-floe distributions according to practical observations, which is used to initialise ice conditions for the CFD&DEM model. Following the completeness of the model, Section 3 provides the prediction and validation of ship resistance in ice floes and shows how the resistance is influenced by ship speed, ice concentration, ice thickness and floe diameter. The CFD&DEM approach is demonstrated to simulate the new shipping scenario with high fidelity and could be used to provide valuable insights for other Arctic engineering purposes.

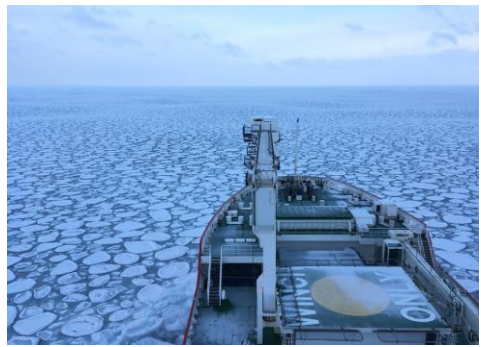

Figure 1: A ship advancing in floating ice floes [36].

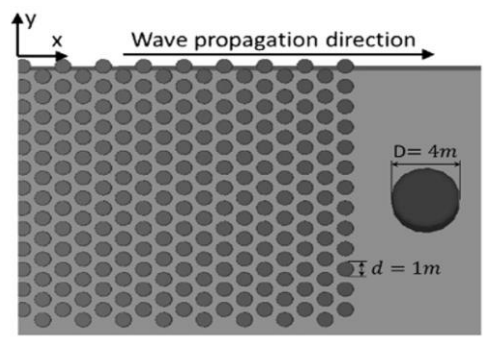

Figure 2: Regular pancake-ice field used by Sun and Shen [34].

## 2. Computational modelling

In order to simulate a ship advancing in floating ice floes, this work performed three steps to build a computational model: (a) CFD, a standard model of ship advancement in open water, where fluid solutions are obtained, including the wake and water resistance of a ship. (b) DEM, for modelling ice floes and their collisions with the ship and nearby floes; those floes obtain fluid force from the CFD solution so that the ship-wave-ice coupling is achieved. (c) floe-distribution algorithms, by which natural ice-floe fields are generated and implemented into the CFD&DEM model. Steps (a) and (b) are based upon the STAR-CCM+ software, and Step (c) is realised using MATLAB scripts.

2.1 CFD ship flow

A modern container ship model, KRISO Container Ship (KCS), was adopted as the ship model for this study. KCS is a typical container ship model employed in computational simulations, and its geometry with appendages is openly accessible [37]. The length of the hull is 230 m at full-scale with a scale ratio of 1:52.667 applied in this study for the purpose of validation, resulting in a model length $L_{pp}$ = 4.367 m and breadth B = 0.611 m. The hull parameters are summarised in Table 1.

Table 1. Main dimensions of the KCS hull.

|  | **Model scale** | **Full scale** |
| --- | --- | --- |
| Length between perpendiculars (m) | 4.367 | 230.0 |
| Waterline breadth (m) | 0.611 | 32.2 |
| Draught midships (m) | 0.205 | 10.8 |
| Trim angle (rad) | 0.0 | 0.0 |
| Block coefficient (-) | 0.651 | 0.651 |
| Wetted surface area (m$^2$) | 3.435 | 8992.0 |

Following the guidelines of the International Towing Tank Conference [38], an open-ocean fluid domain was built with the recommended domain size and boundary conditions, as shown in Figure 3. The computational domain is three-dimensional, defined by the earth-fixed Cartesian coordinate system O-*xyz*. The (*x, y*) plane is parallel to the horizon, and the *z*-axis is positive upwards. The domain size is sufficiently large to avoid the ship-generated waves being reflected from the boundaries. The lower part of the domain is filled with water and the remainder is filled with air. The hull is fixed at the free surface according to its design draught and the hull surface is modelled as a no-slip wall. The water was initialised as flowing with a constant velocity ($U_{water}$) against the bow of the hull, and a constant velocity condition was applied to the inlet boundary to maintain a stable water flow entering the domain. Thus, a relative velocity exists between the ship and water, where $U_{water}$ indicates the advancing speed of the ship in calm water. The ship speed can be converted to Froude number $Fr = U/\sqrt{g \times L_{pp}}$, where $g$ and $L_{pp}$ are gravitational acceleration and ship length respectively. The fixed-dynamic-pressure condition is applied to the outlet, and the zero-gradient condition is applied to other boundaries.

The solution of the fluid domain was obtained by solving the Reynolds-averaged Navier-Stokes (RANS) equations for an incompressible Newtonian fluid:

$$\nabla \cdot \overline{\mathbf{v}} = 0 \tag{1}$$

$$\frac{\partial(\rho \overline{\mathbf{v}})}{\partial t} + \nabla \cdot (\rho \overline{\mathbf{v}\mathbf{v}}) = -\nabla \overline{p} + \nabla \cdot \left(\overline{\tau} - \rho \overline{\mathbf{v}'\mathbf{v}'}\right) + \rho g \tag{2}$$

where $\overline{\mathbf{v}}$ is the time-averaged velocity, $\mathbf{v}'$ is the velocity fluctuation, $\rho$ is the fluid density, $\overline{p}$ denotes the time-averaged pressure, $\overline{\tau} = \mu[\nabla v + (\nabla v)^T]$ is the viscous stress term, μ is the dynamic viscosity and $g$ is gravitational acceleration set at 9.81 m/s². Since the RANS equations have considered the turbulent fluid, the Shear Stress Transport (SST) k − ω model [39] was adopted to close the equations. The SST k − ω model has been proposed to be the most appropriate option among standard RANS turbulence models for predicting the flow field around a ship hull [40].

The free surface between the air and water was modelled by the Volume of Fluid (VOF) method [41]. The VOF method introduces a passive scalar α, denoting the fractional volume of a cell occupied by a specific phase. In this case, a value of α = 1 corresponds to a cell full of water and a value of α = 0 indicates a cell full of air. Thus, the free surface, which is a mix of these two phases, is formed by the cells with 0 < α < 1. The elevation of the free surface along time is obtained by the advection equation of α, expressed as Equation (3). For a cell containing both air and water, its density and viscosity are determined by a linear average according to Equation (4) and Equation (5). In this study, $\rho_{water}$ = 998.8 kg/m³, $\mu_{water}$ = 8.90×10⁻⁴ N·s/m²; $\rho_{air}$ = 1 kg/m³, $\mu_{air}$ = 1.48×10⁻⁵ N·s/m². The governing equations of the fluid domain were discretised and solved using the Finite Volume method [42]; Figure 4 shows the mesh layout of the model, in which high resolutions are applied to the regions where ship-wave-ice interactions are expected to happen, resulting in a total cell number of around three million.

$$\frac{\partial \alpha}{\partial t} + \nabla \cdot (\overline{\mathbf{v}} \alpha) = 0 \tag{3}$$

$$\rho = \alpha \rho_{water} + (1-\alpha) \rho_{air} \tag{4}$$

$$\mu = \alpha \mu_{water} + (1-\alpha) \mu_{air} \tag{5}$$

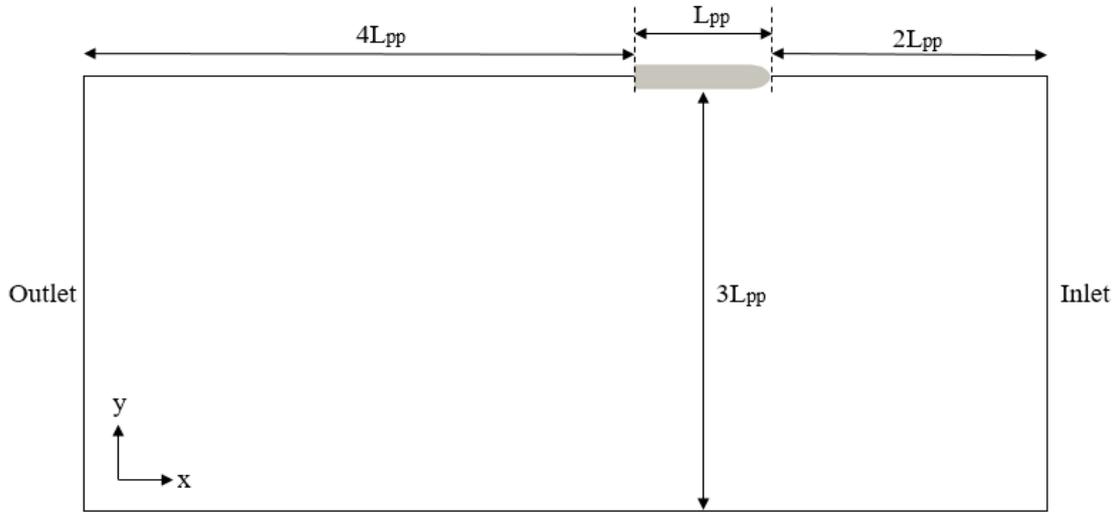

(a) Plan view; only half of the domain is shown but no symmetry plane condition is applied

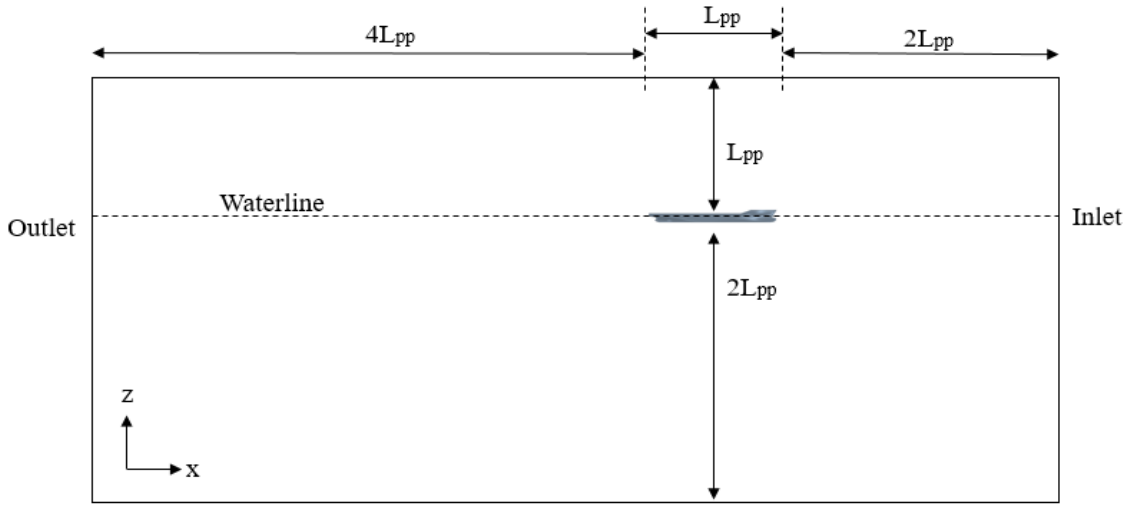

(b) Profile view

Figure 3: Illustration of the computational domain with dimensions.

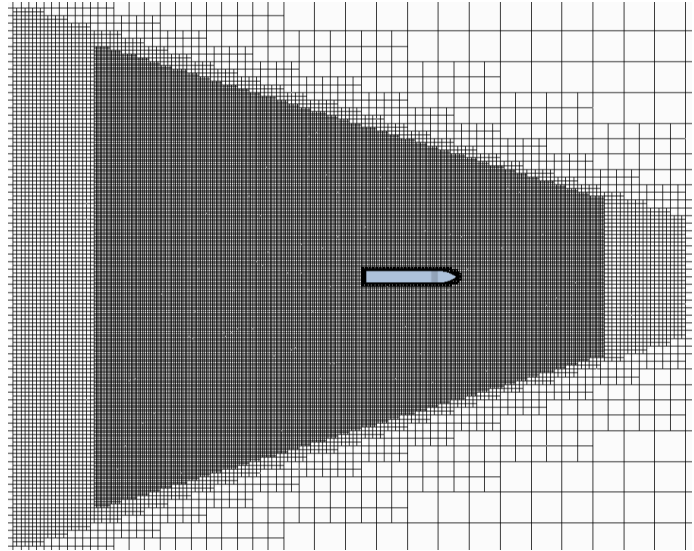

(a) Plan view

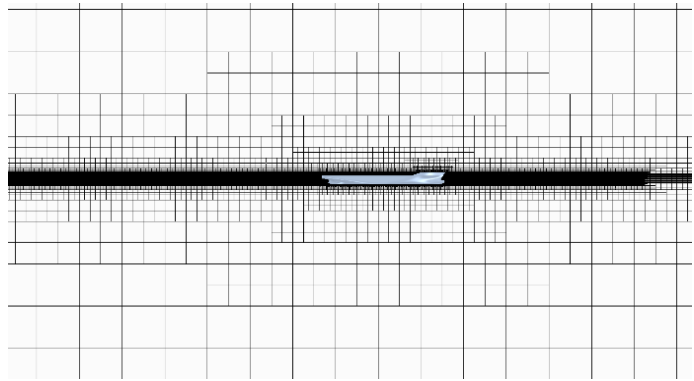

(b) Profile view

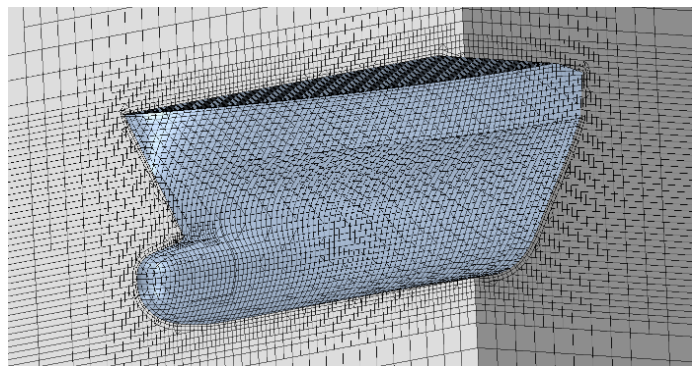

(c) Isometric view of bow region

Figure 4: Mesh layout of the model, in which local refinements are applied at the Kelvin-wave region, the free surface region and around the hull geometry.

## 2.2 DEM ice floes

Importing the ice floes is achieved by a novel array-injection method. The simulation first runs for a certain time without sea ice to allow the fluid domain to achieve a steady-state, i.e. when the ship wake become stable. Subsequently, an array of ice floes is injected into the computational domain near the inlet boundary, as shown in Figure 5(a). The ice floes are initialised to have the same velocity as the water flow ($U_{ice} = U_{water}$), and one ice floe array is injected to the same region every $t_{inject} = L_{array}/U_{ice}$, so that the next ice floe array can just follow the former one, as shown in Figure 5(b). Thus, the injection of ice floes does not influence the stability of solutions around the ship, and the ship can keep advancing in a continuous ice-floe zone, as desired, shown in Figure 6. With this method, an ice-floe route of unlimited length can be achieved without the need for a very long domain, which significantly saves computational costs.

The ice floes are modelled as DEM particles in the Lagrangian framework moving in the Eulerian fluid domain [43]. Each ice floe is modelled as a rigid thin disk, as ice deformation is negligible in the present problem [44]. The density of ice is set at $\rho_{ice}$ = 900 kg/m$^3$, while ice diameter $D$ and thickness $h$ are variable. The movement of an ice floe can be considered as the combination of translation and rotation, which was solved with the rigid-body motion equations in the body-fixed system based on the mass centre of the floe $G-x'y'z'$:

$$\mathbf{F} = m \frac{d\vec{V_G}}{dt} \tag{7}$$

$$\mathbf{T} = [J] \cdot \frac{d\vec{\omega_G}}{dt} + \vec{\omega_G} \times ([J] \cdot \vec{\omega_G}) \tag{8}$$

where $\mathbf{F}$ and $\mathbf{T}$ are the total force and torque on the ice floe, induced by the gravity, the hydraulic load from the surrounding fluid $\mathbf{F_h}$ and the contact force $\mathbf{Fc}$ from ship-ice contact and ice-ice contact; $m$ and $[J]$ are the mass and inertia moment tensor respectively, and $V_G$ and $\omega_G$ are the translational and rotational velocity vectors of the ice floe respectively.

The hydraulic force $\mathbf{F_h}$ can be calculated based on the solution from the fluid domain, as expressed in Equation (9). Specifically, each DEM ice floe is projected into the fluid mesh, so that the cells around the floating disk are partially blocked by the disk volume and fluid solutions are interpolated on the disk surface, i.e. the fluid solution is integrated over the surface of each ice floe to contribute the total force and torque. Therefore, the cells around the free-surface area should be small enough to capture the floe geometry; as in Figure 4, around 4 cells per ice thickness is applied in this study, and this needs to be refined accordingly when thinner ice is applied. The timestep size used in CFD is 0.001 s so that the Courant number is less than one, while each CFD timestep is split into 100 DEM sub-timesteps so that the collisions can be sufficiently solved, which shows DEM dominates the time requested for the CFD+DEM computation.

$$\boldsymbol{F_h} = \int (-\bar{p}\,\boldsymbol{n} + \bar{\tau} \cdot \boldsymbol{n})\,dS \tag{9}$$

The CFD+DEM is a one-way coupling mechanism, i.e. the ice movement does not provide feedback to the fluid domain thus the wave radiation is not influenced by floes. Two-way coupling is out of the scope of this paper since it will induce very high computational costs. In addition, the wave radiation mainly influences the floes moving away from the ship, so it is deemed to have little influence on the ship performance, as far as the ice concentration is not high enough to produce a considerable wave reflection. The validation in Section 3.1 shows the one-way mechanism is enough to provide accurate resistance predictions when ice concentration is up to 70%.

The contact force $\mathbf{F_c}$ is calculated by a penalty method [45], where ship/ice and ice/ice are allowed to have an overlap, according to the motion solutions. The overlap is modelled as a linear spring-dashpot system where the spring ($k$) accounts for the elastic response and the dashpot ($\eta$) reflects the energy dissipation during the contact, by which the normal and tangential components of $\mathbf{F_c}$ are calculated according to Equation (10) and (11) respectively. Subsequently, the contact force pushes the overlapped bodies apart so that the overlap is minimised in the final solution.

$$\mathbf{F_n} = -kd_n - \eta v_n \tag{10}$$

$$\mathbf{F_t} = \begin{cases} -kd_t - \eta v_t, & \text{if } |d_t| < |d_n|C_f \\ |kd_n|C_f \cdot \boldsymbol{n}, & \text{if } |d_t| \geq |d_n|C_f \end{cases} \tag{11}$$

where $d_n$ and $d_t$ are overlap distances in the normal and tangential directions respectively, $v_n$ and $v_t$ are the normal and tangential components of the relative velocity between two contact bodies, $C_f$ is the friction coefficient, which is set at 0.35 for ice-ice contact and 0.05 for ship-ice contact; $k$ is set at $6\times10^4$ N/m and $\eta = 2C_{damp}\sqrt{kM_{eq}}$, in which $C_{damp}$ is set at 0.067 and $M_{eq}$ is the equivalent mass of two contact bodies, calculated as $M_{eq} = M_A M_B/(M_A + M_B)$. These parameter values are selected based on [11,46], and a slight adjustment was applied following a comparison with experimental results.

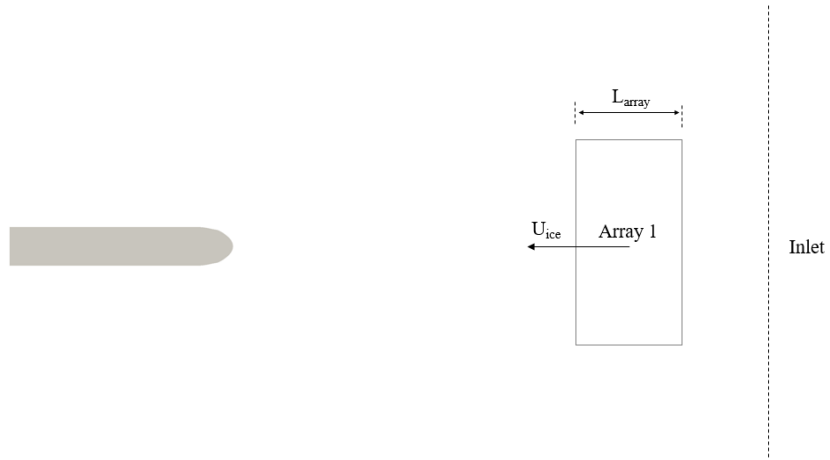

(a) t = $t_1$ when the first ice array is injected

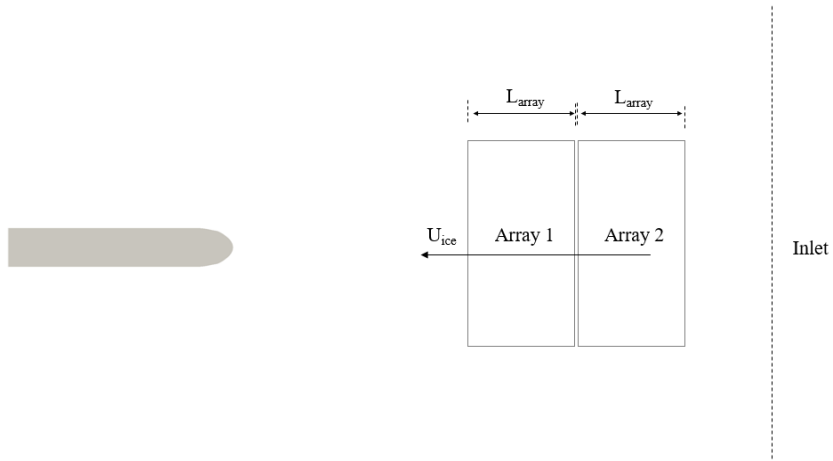

(b) t = $t_1$ + $t_{inject}$ when the second ice array is injected

Figure 5: Illustration of how ice floes are imported.

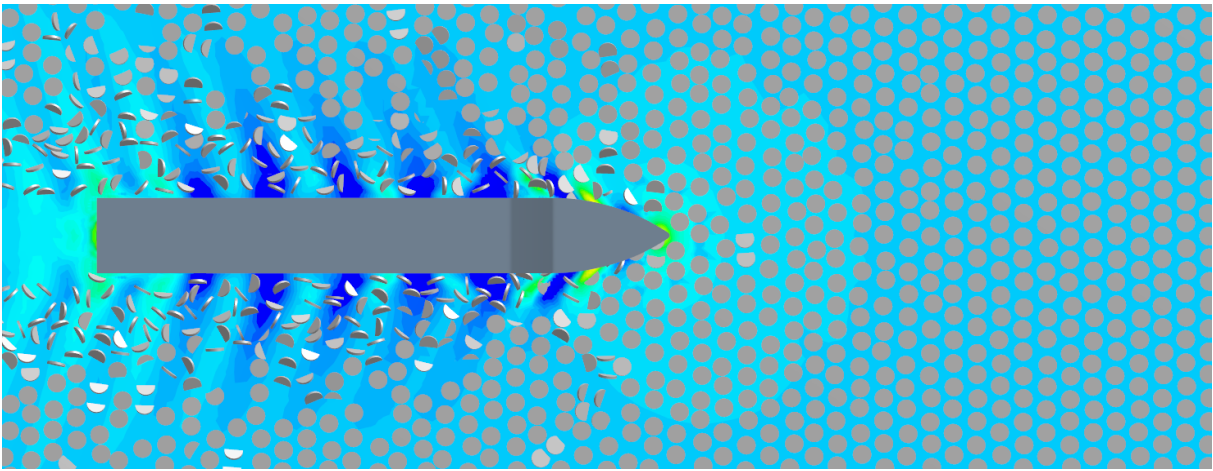

Figure 6: Simulation view of a ship advancing in floating ice floes.

2.3 Floe-distribution algorithms

The ice-floe array initialises floes to be of uniform size and equidistant to each other, as shown in Figure 6, which is not a realistic condition and introduces artificial regularity into the ice load. For example, such setup makes ship-ice contacts always occur at the same locations on the hull, which dictates the resistance as well as structural response of the ship. This defect motivates the development of appropriate Floe-Size-Distribution (FSD) algorithms to introduce ice floes into a computational model where floe size and location can follow natural conditions observed in polar regions.

There are two principal features for realistic FSD: (a) ice floes are a mixture of different sizes, and (b) the location of each ice floe should be randomly distributed. The distribution of floe size against possibility is suggested to follow a log-normal function [4], an example is shown in Appendix Figure 1. For this purpose, two algorithms have been developed to generate a natural ice-floe map within a desirable domain. Main inputs of both algorithms are the length and width of the domain, as well as the ice concentration (C), defined as the ice-covered area divided by the whole sea-surface area, i.e. area of the domain. There are two constraints for the proposed problem: the first one is to avoid overlapping between ice floes, and the other is to assure that the centre of every floe lays inside the domain boundary. In this study, the thickness of all floes is assumed to be consistent.

Algorithm (I) first calculates the number of ice floes and marks them from 1 to $n_p$, and then it randomly injects all ice floes inside the domain and their initial locations are stored in a location matrix. Subsequently, the algorithm proceeds by considering every piece separately, starting from i=1 to i=$n_p$. The algorithm first checks if overlapping occurs (with other ice pieces and with the boundary of the domain). Where an overlapping occurs, the vicinity of the ice piece is searched to see whether there is a sufficient space for it to move over (direction vector revolves 360 degrees and searches the space in the vicinity of the original position). If there is sufficient space in the vicinity, the ice floe will be moved there, and the location matrix is updated. If there is not enough available space in the vicinity, the ice piece is then randomly reallocated to another position. The process iterates until all ice pieces are settled and no constraints are violated, depicted in Appendix Figure 2.

Algorithm (I) is able to rapidly yield ice distributions up to C = 60%. Nevertheless, it gets slower for higher C values, since this means there is less open-water space in the domain and thus it is harder to move the floes into free space. Therefore, in order to deal with high-concentration ice conditions, Algorithm (II) has been developed. It is based on the Genetic Algorithm (GA) [47], which is an optimisation technique inspired by Darwin's principles of the survival of fittest individuals. Algorithm (II) defines a penalty factor to indicate the overlapping between ice floes, where a higher penalty value corresponds to a higher overlapping area. The target penalty factor is set to be zero in this study, meaning no overlapping at all. In other words, the target penalty factor can be set to be larger than zero when certain ice overlapping is allowed, which is suitable for modelling ice herding and rafting [13,48].

To initialise Algorithm (II), a certain number ($M_0$) of parent distributions are randomly obtained, corresponding to $M_0$ penalty factors. Subsequently, the algorithm selects a small number of solutions ($M_1$) with the lowest penalty factors, which will be used as parents to produce child distributions, yielding the same overall amount of $M_0$ solutions for the next generation. Namely, only fittest individuals survive, while the rest of them are discarded by the algorithm. Thus, the average of selected solutions is better than the previous ones, leading to a higher probability to get better child solutions. In addition, $M_2$ best solutions, known as elite individuals, automatically survive to the next generation without a change; while the rest of the solutions ($M_1 - M_2$) relocate ice floes to get ($M_0 - M_2$) new distributions, which are combined with the elite individuals to form the next parents. Iterations are performed until one distribution has been found to achieve the target penalty, as illustrated in Appendix Figure 3. In this study, $M_0 = 2000$, $M_1 = 1000$ and $M_2 = 100$.

Algorithm (II) is slower compared to (I) to obtain a desired solution, while it is capable of generating high-concentration floe fields. The output from both algorithms is a matrix listing the *x-y* coordinates of every ice piece and the corresponding floe size, while *z* coordinate is aligned to the buoyancy-gravity equilibration of each floe. Figure 7 shows two samples of floe distributions obtained by algorithms (I) and (II) respectively, where the targeted FSD and C are perfectly achieved. Although the shape of ice floes was set to be circular to fit the pancake-ice nature, both generators are capable of modelling other ice shapes. Moreover, the distribution law can be easily switched to any field measurements or aerial observations, such as [3,49].

For a specific FSD and C condition, the algorithms can generate the corresponding ice field for a ship to enter, which is used to replace a regular array as in Figure 6. Updated simulations are shown in Figure 8, in which the artificial regularity of ice load has been avoided. Up to this point, the computational modelling for the present work is completed. Noting that the introduced algorithms are not only compatible with current work, they are also useful for other polar modelling involving ice floes.

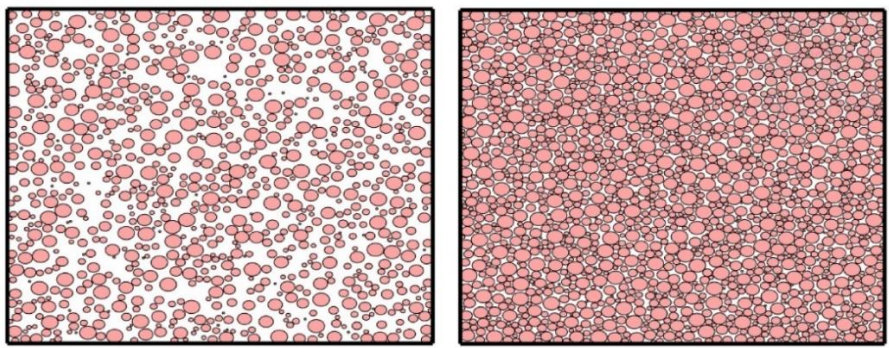

Figure 7: Ice-floe fields obtained by Algorithm (I), C = 40% (left side); and by Algorithm (II), C = 70% (right side).

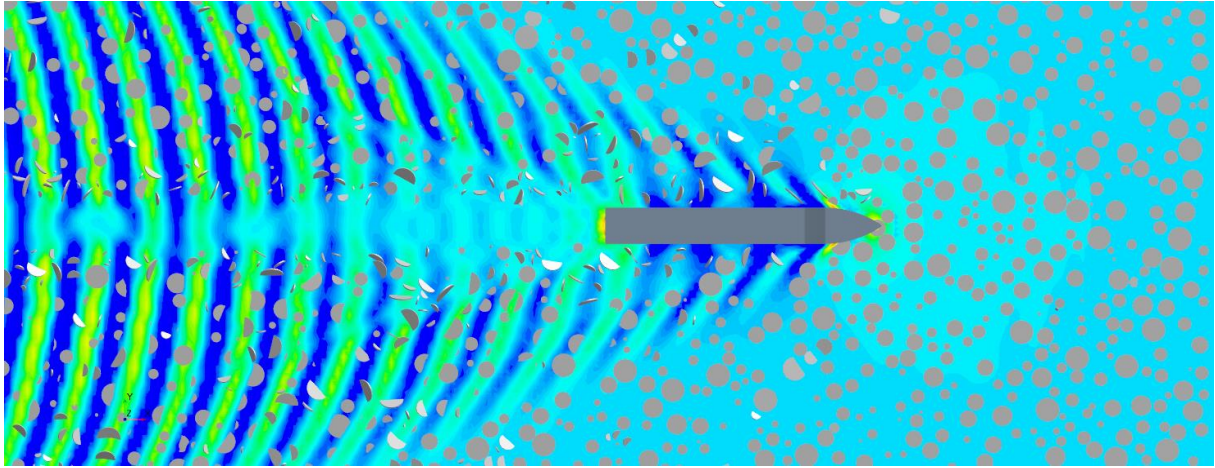

(a) Ice concentration = 30%

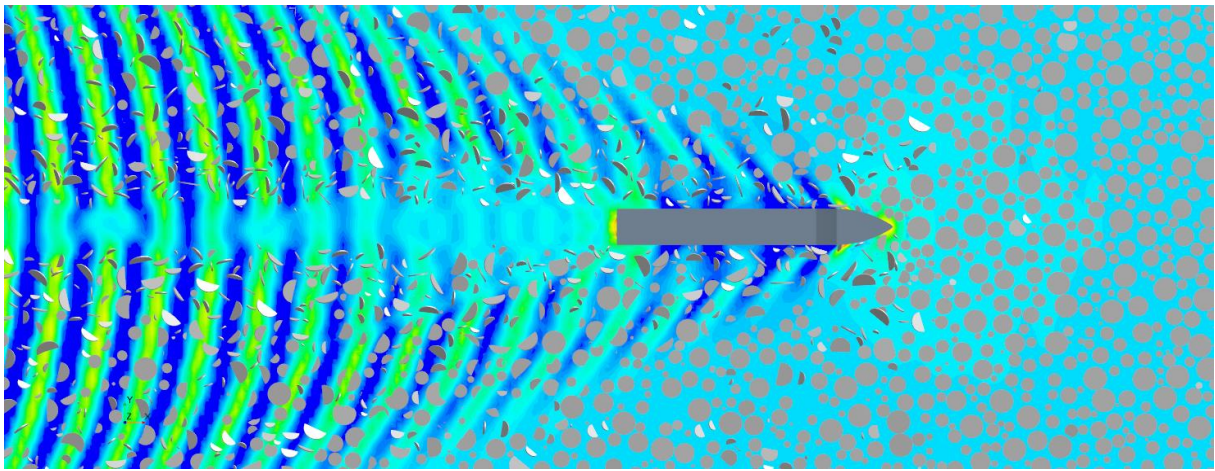

(b) Ice concentration = 50%

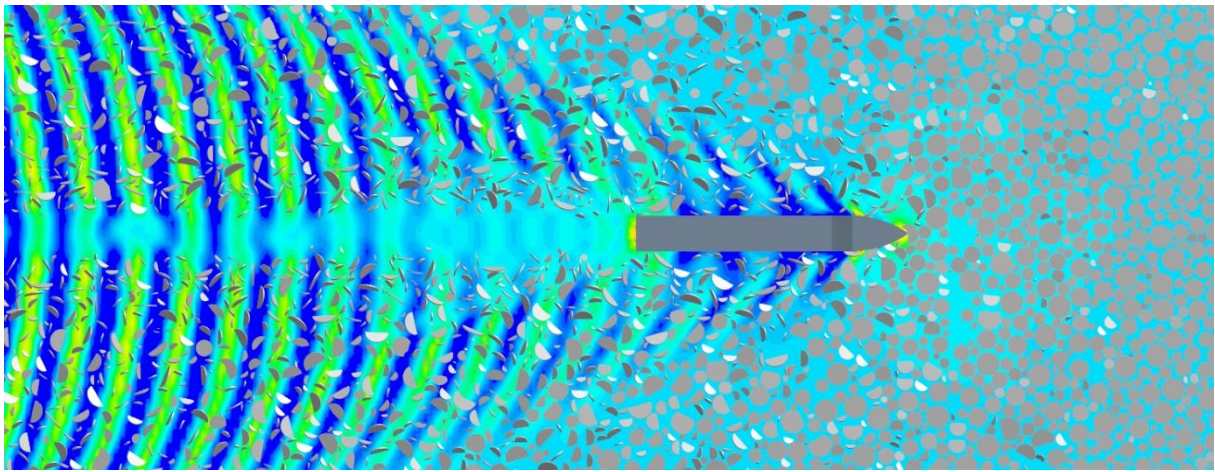

(c) Ice concentration = 70%

Figure 8: Simulation view of a ship advancing in floating ice floes, with natural floe fields implemented.

## 3. Ship resistance

The predicted steady-state simulations are shown in Figure 8: when a ship is advancing in ice floes, ship-ice collisions occur at the bow area, causing the floes to be pushed aside and rotate within the wake, and in some cases, floes can slide along the ship before being pushed away. This work focuses on the ship resistance, namely the force born by the ship in its advancement direction, since this determines the power required for such shipping. The total resistance of the ship consists of an ice resistance $R_{ice}$ induced by the floes and a water resistance $R_{water}$ similar to an open-ocean case ($R_{total} = R_{ice} + R_{water}$). $R_{ice}$ occurs due to ship-ice collisions at the bow area as well as the force due to floes sliding along the sides of the ship. It is calculated by summing of ship-ice contact impulses in the *x*-direction over a certain time period and then dividing the total impulse by the time. An example of the impulse time-series is shown in Figure 9, where the varying values correspond to ship contacts with floes of different sizes. Randomness can be seen in the time-series, which is reasonable due to the nature of the floe-distribution algorithms. Therefore, before taking a resistance result, each simulation needs to be run for a sufficiently long time (more than 40 s) to offset this randomness (until $R_{ice}$ does not noticeably change with runtime, as shown in Figure 10). Every 40 seconds of the CFD&DEM simulation costs approximately two days in real time using a 100-core High Performance Computation (Intel Xeon E5-2630v3) provided by University College London (UCL).

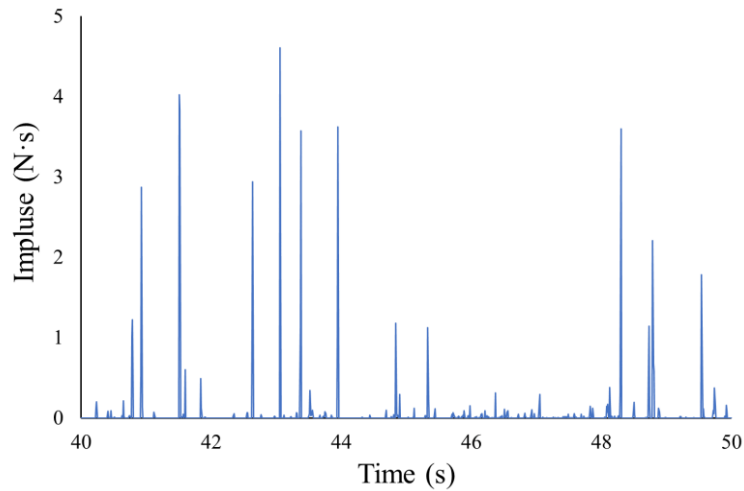

Figure 9: Time series of ice-induced impulse against the direction of ship advancement (obtained when Fr = 0.12, C = 60%, *h* = 0.02 m and FSD follows [4]).

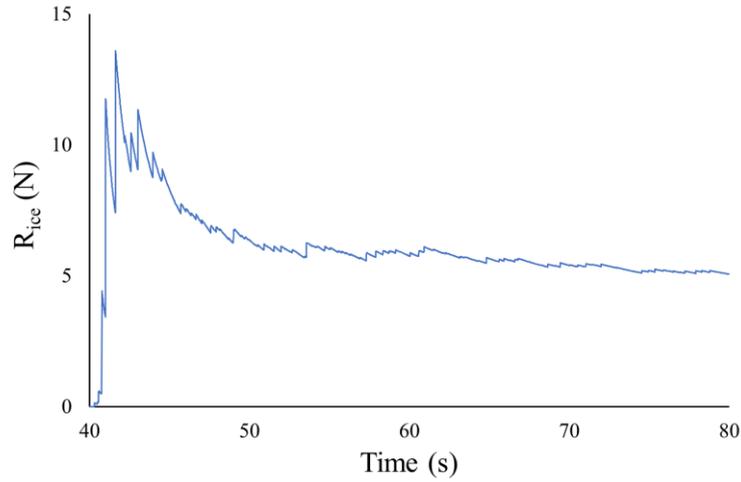

Figure 10: Time series of ice resistance: an oscillation is shown when the ship just enters the floe field, and $R_{ice}$ gradually stabilises with runtime.

3.1 Validation

The predicted resistance has been validated against the experiments conducted in the towing tank of Harbin Engineering University [4]. The computational resistance is compared with available experimental data, where ship speeds ranged from Froude number Fr = 0.06 to 0.18, and ice concentration C = 60% and 70%, ice thickness $h$ = 0.02 m. The comparison between computational and experimental results is presented in Figure 11, and the good agreement indicates a reliable accuracy of the model in predicting ship resistance in ice floes. Figure 11 also shows the open water component, and $R_{ice}$ can be seen as the difference between the total-resistance curve and the water-resistance curve. This work only considers ice concentrations of up to 70%, since ice rafting may occur with higher concentrations (when C > 79% according to the result of Hopkins and Tuhkuri [11]), while the capability of introduced CFD&DEM approach on modelling ice rafting needs further verifications, where the floe contact is between surfaces rather than edges.

3.2 Ship speed and ice concentration

Extended simulations were performed to investigate the influence of ship speed and ice concentration. Figure 12 presents the ice resistance for varying ship speeds and ice concentrations. Figure 12(a) shows the variation trend of the resistance; regression analyses indicate the increasing powers of ship speed and ice concentration are respectively around 1.2 and 1.5 in the examined range. Thus, the increase of ice resistance with an increasing ship speed is slower than that of open water resistance whose power is recognised to be 2. Figure 13(b) presents the ratio of $R_{ice}$ divided by corresponding $R_{water}$, which shows $R_{ice}$ is more influential when the ship is relatively slow and such an effect is more pronounced for high ice concentrations. This is because a faster ship generates larger waves, which tends to push the floes away thus reducing the ship-ice contact. This corroborates the importance of the inclusion of fluid flow in the present work, as waves can significantly influence the ship-ice interaction and ice resistance. Nevertheless, it should be noted the simulations are conducted at model scale; in full scale the floes may not be pushed away to the same extent: the mass of the floes scales as $D^3$, but the forces due to waves are known as scaled by $D^2$, where D is floe diameter.

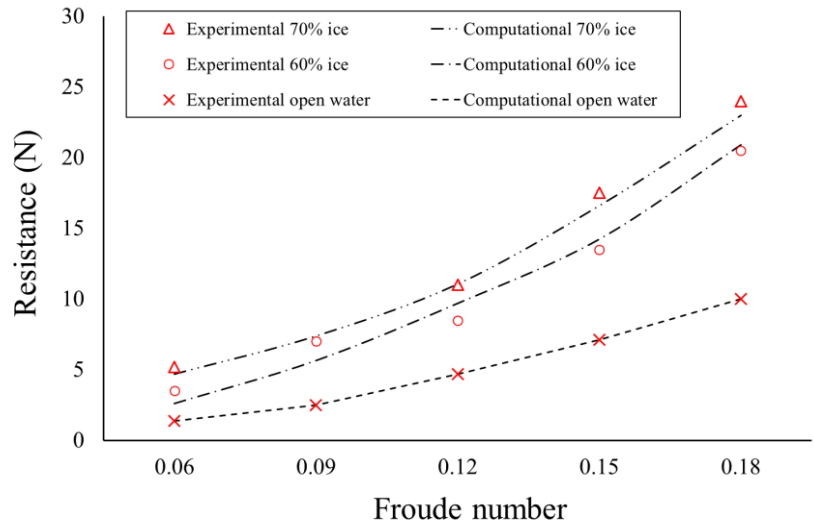

Figure 11: Experimental [4] and computational total resistance of model-scale KCS operating in ice concentration 60% and 70%, alongside the water component.

3.3 Floe diameter and ice thickness

The influence of floe diameter is analysed by globally scaling all floes, i.e. multiplying the FSD curve with a factor. Figure 13 compares varying floe-diameter conditions while ice concentration remains constant. It can be seen that larger floe diameters lead to sparser ice (fewer floes) and lower the collision frequency, while significantly increasing the peak impulses. The overall ice resistance with different floe diameters are plotted in Figure 14, in which the resistance obviously increases with increased floe diameters; despite a lower collision frequency, the force integration over time still increases due to the peak values. This indicates that, with the same ice concentration, the effect of floe diameter on ship resistance is more dominant than that of collision frequency. Overall, $R_{ice}$ reveals a linear trend with varying floe diameter.

The influence of ice thickness on the resistance is studied by varying $h$ while keeping other parameters constant. Figure 15 presents the ice resistance when $h = 0.004 - 0.02$ m, and for C = 40% and 60%. Similar to floe diameter, $R_{ice}$ reveals a linear trend with varying ice thickness. When FSD and C are held constant, varying $h$ can also be considered as varying the mass of each floe, thus total ice mass ($\sum m_{ice}$) is varied according; similarly, Gong et al. [17] reported a quasi-linear relationship between $\sum m_{ice}$ pushed by a ship and the corresponding resistance. However, in Figure 12 where total ice mass increases with the square of ice concentration, and in Figure 13 and 14 where total ice mass remains constant when floe diameters are changed, the linear relationship between $\sum m_{ice}$ and ship resistance does not exist. This indicates it cannot conclude ship resistance varies linearly with $\sum m_{ice}$ in this ice-floe scenario, since waves and FSD have introduced high complexities to the associated physics. Each variable needs to be analysed separately in this case.

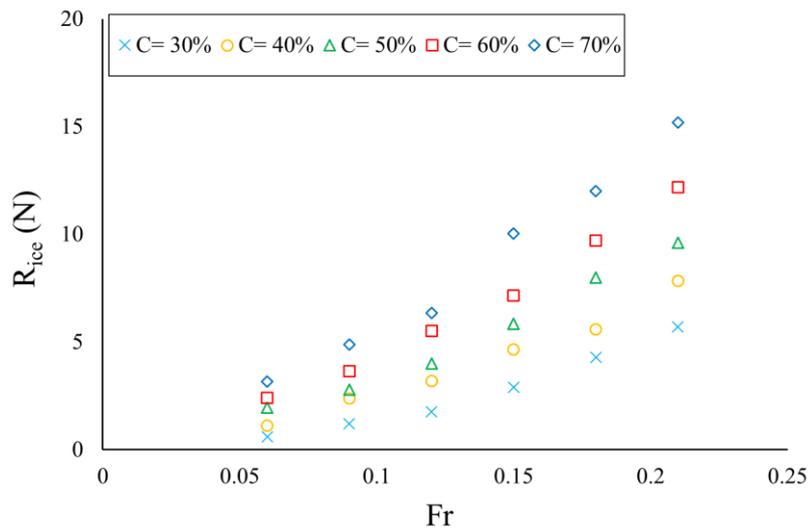

(a) Ice resistance values

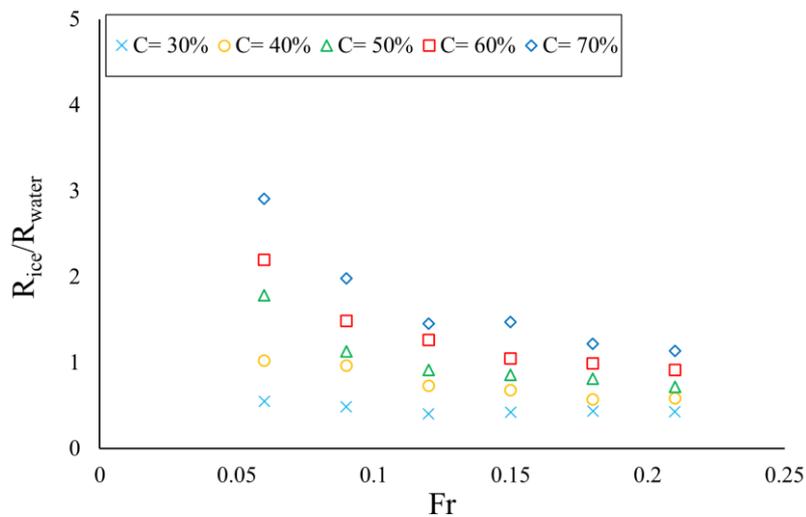

(b) Ice resistance normalised by water resistance

Figure 12: Ice-floe resistance in varying ice concentrations and ship speeds.

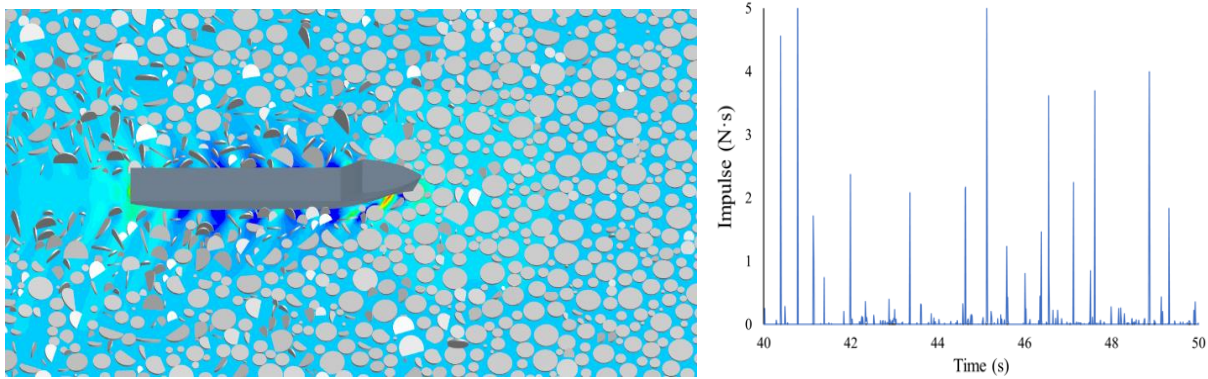

(a) Floe diameters are the same as those in [4]

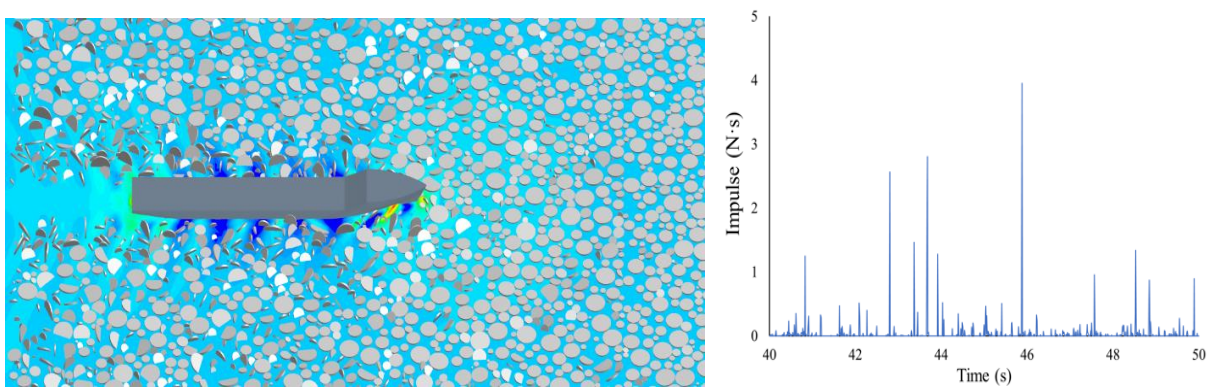

(b) Floe diameters are 60% of those in [4]

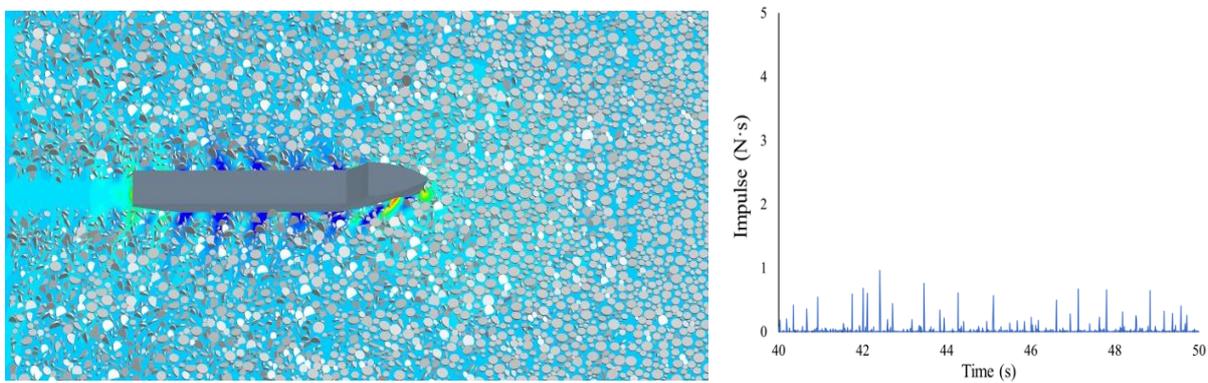

(c) Floe diameters are 20% of those in [4]

Figure 13: Ship advancing in different-sized floes (left side) and corresponding time-series of resistance impulse (right side); obtained when Fr = 0.15, $h$ = 0.02 m and C = 60%.

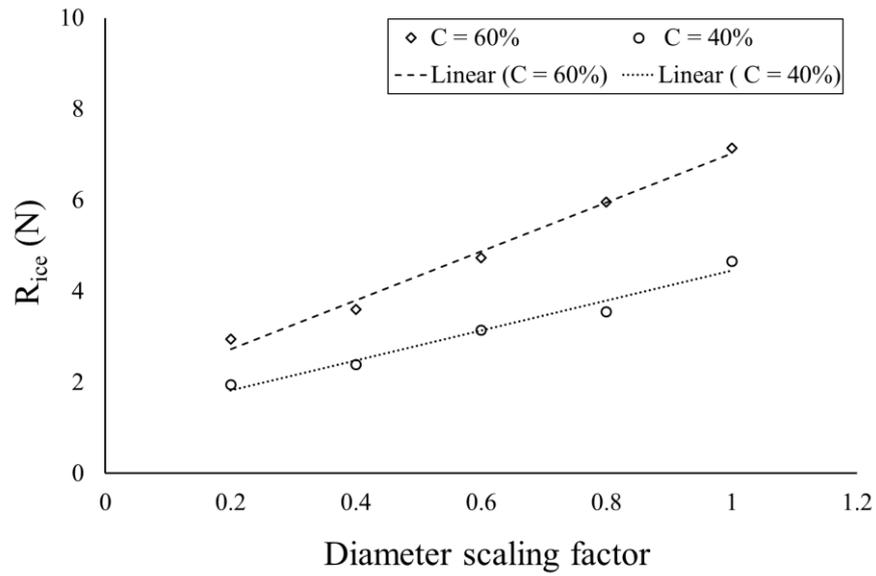

Figure 14: Ice-floe resistance in different-sized floes (floe diameters of [4] globally scaled by a factor), obtained when Fr = 0.15 and $h$ = 0.02 m.

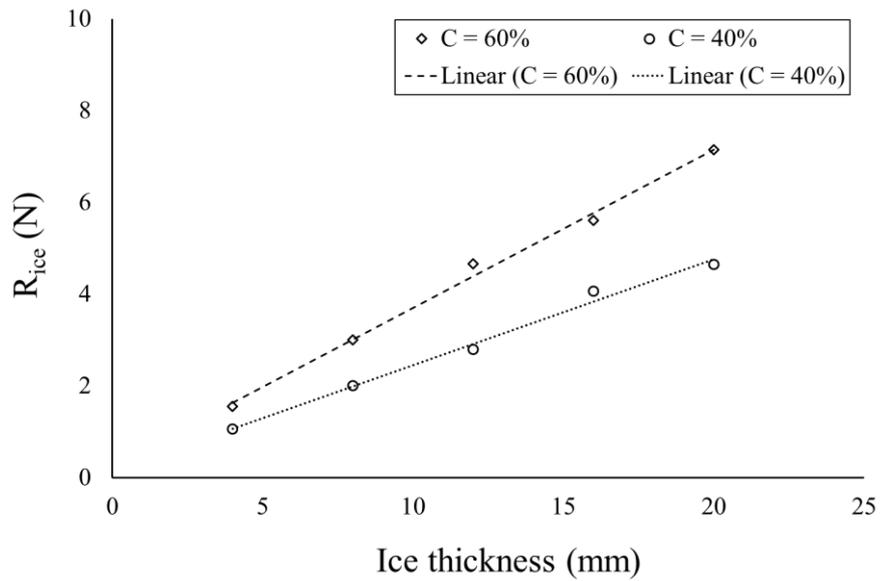

Figure 15: Ice-floe resistance in varying ice thicknesses, obtained when Fr = 0.18.

## 4. Conclusions

A CFD&DEM approach has been introduced simulate a ship advancing in floating ice floes, since such a condition has been predicted to be the main navigation environment of future Arctic shipping. Relevant numerical theories and practicalities have been presented in detail: CFD is incorporated with DEM to model desirable ship-wave-ice interactions, and two algorithms for generating natural ice-floe fields have been provided. The proposed model shows the capability to simulate and analyse the proposed problem with high fidelity, and it is validated to be accurate in predicting the ship resistance induced by ice floes.

This work presents the first model that includes CFD flow to account for waves in the ship-wave-ice interaction process, which proves to be of great importance: the ship-generated waves have been demonstrated to reduce the ship-ice collision, and it leads to the finding that the ice-added resistance is more influential when the ship is relatively slow. Nevertheless, future work should examine the CFD&DEM approach in modelling ice rafting, which might influence the interaction process when the ice concentration is greater than 70% [11]; thus, the predictions of this study are below this limitation.

A series of simulations have been performed to investigate how the ice resistance is influenced by ship speed, ice concentration, ice thickness and floe diameter. It was found that ship speed and ice concentration govern the resistance with a power of 1.2 and 1.5 respectively, whilst floe diameter and ice thickness both show a linear effect. This will be useful to derive empirical equations to provide rapid estimates for ship resistance in this emerging scenario - floating ice floes.

**Acknowledgements**

This work is part of a project that has received funding from the European Union's Horizon 2020 research and innovation programme under grant agreement No 723526 - SEDNA: Safe maritime operations under extreme conditions; the Arctic case. The first author is grateful to Lloyds Register Foundation, UCL Faculty of Engineering Science and China Scholarship Council, for providing his PhD scholarship; he appreciates Professor Guoxiong Wu who supports this funding.

**Appendix**

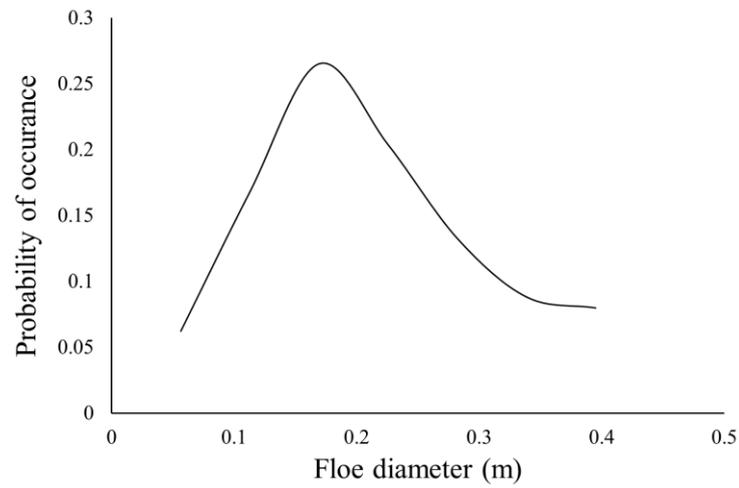

Appendix Figure 1: FSD used by Guo et al. [4].

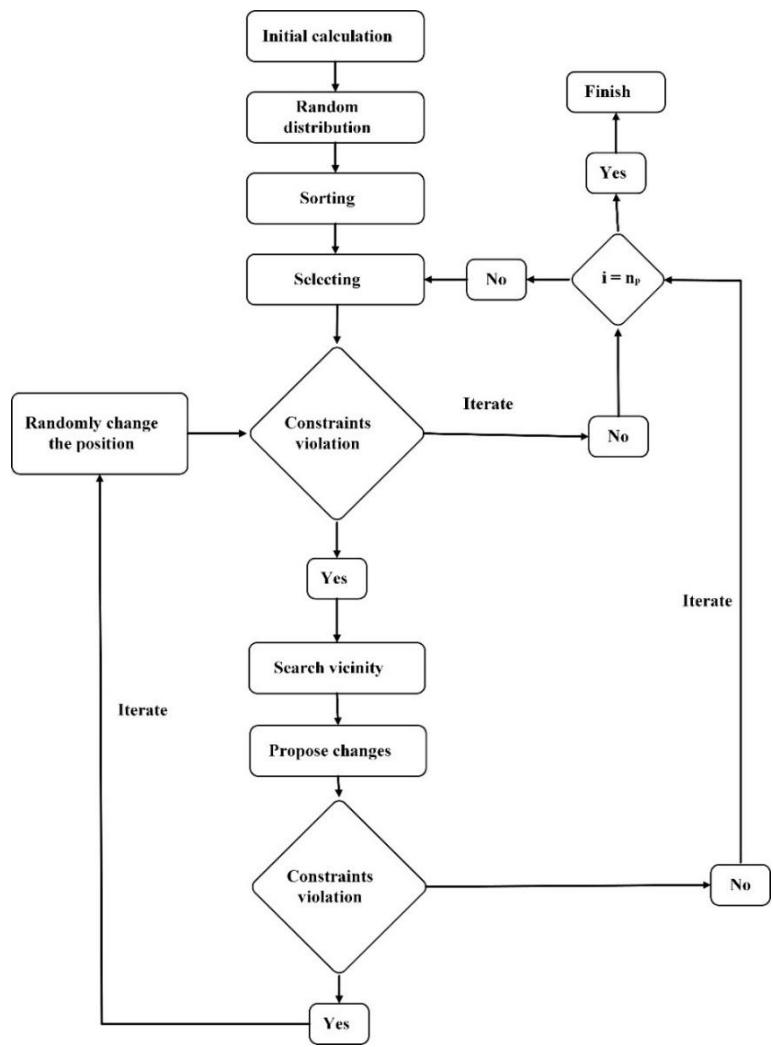

Appendix Figure 2: Flow chart of Algorithm I.

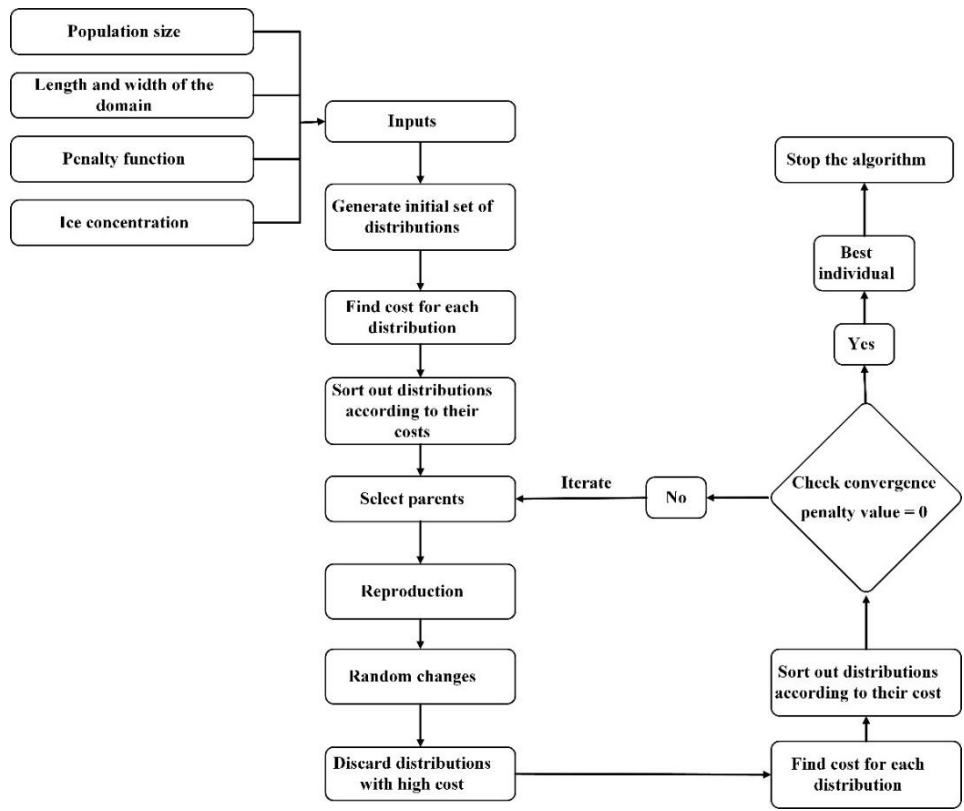

Appendix Figure 3: Flow chart of Algorithm II.